\documentclass[twocolumn,prl,showpacs,floatfix,superscriptaddress]{revtex4-2}
\usepackage{graphicx}
\usepackage{stickstootext}
\usepackage{amsthm}
\usepackage[stickstoo,vvarbb]{newtxmath}
\usepackage{microtype}
\usepackage[colorlinks=true, linkcolor=blue, anchorcolor=blue, citecolor=blue, filecolor=blue, urlcolor=blue]{hyperref}
\newcommand{\ikf}{Institut f\"ur Kernphysik, J. W. Goethe Universit\"at, Max-von-Laue-Str. 1, 60438 Frankfurt, Germany}
\newcommand{\esrf}{ESRF, 6 Rue J. Horowitz, BP 220, 38043 Grenoble Cedex 9, France}

\begin{document}
% \draft command makes pacs numbers print
%\draft
\title {
\large
Photon-momentum-induced molecular dynamics \\
in photoionization of N$\boldsymbol{_2}$ at $\boldsymbol{h\nu=40}$~keV}
% repeat the \author\address pair as needed
\author{M.~Kircher} \address{\ikf}
\author{J.~Rist} \address{\ikf}
\author{F.~Trinter}
\address{
	FS-PETRA-S, Deutsches Elektronen-Sychrotron (DESY), Notkestr. 85, 22607 Hamburg, Germany}
\address{
	Molecular Physics, Fritz-Haber-Institut der Max-Planck-Gesellschaft, Faradayweg 4, 14195 Berlin, Germany
}
\author{S.~Grundmann} \address{\ikf}
\author{M.~Waitz} \address{\ikf}
\author{N.~Melzer} \address{\ikf}
\author{I.~Vela-Perez} \address{\ikf}
\author{T.~Mletzko} \address{\ikf}
\author{A.~Pier} \address{\ikf}
\author{N.~Strenger} \address{\ikf}
\author{J.~Siebert} \address{\ikf}
\author{R.~Janssen} \address{\ikf}
\author{V.~Honkim\"aki} \address{\esrf}
\author{J.~Drnec} \address{\esrf}
\author{Ph. V. Demekhin}
\address{
	\mbox{Institut f\"ur Physik und CINSaT, Universit\"at Kassel, Heinrich-Plett-Str. 40, 34132 Kassel, Germany}
}
\author{L.~Ph.~H.~Schmidt} \address{\ikf}
\author{M.~S.~Sch\"offler} \address{\ikf}
\author{T.~Jahnke} \address{\ikf}
\author{R.~D\"orner}
\email{doerner@atom.uni-frankfurt.de}
\address{\ikf}

\begin{abstract}
We investigate K-shell ionization of N$_2$ at 40~keV photon energy. Using a COLTRIMS reaction microscope we determine the vector momenta of the photoelectron, the Auger electron and both N$^+$ fragments. These fully differential data show that the dissociation process of the N$_2^{2+}$ ion is significantly modified not only by the recoil momentum of the photoelectron, but also by the photon momentum and the momentum of the emitted Auger electron. We find that the recoil energy introduced by the photon and the photoelectron momentum is partitioned with a ratio of approximately $30/70$ between the Auger electron and fragment ion kinetic energies, respectively. We also observe that the photon momentum induces an additional rotation of the molecular ion.
\end{abstract}
\maketitle

The linear momentum of the photon is one of its few fundamental properties. It is at the heart of many elementary effects of light-matter interaction. Prominent examples are ionization by Compton scattering and laser cooling. In case of photoionization, however, it is the role of the photon's energy which is typically considered predominantly, and effects caused by the photon’s momentum are to a large extent neglected. Nonetheless, the photon momentum influences photoionization in two ways: Firstly, momentum conservation dictates that in every individual photoionization event the photon momentum is transferred to the center of mass of the photofragments, i.e. the electron(s) and the ion(s). Due to the mass ratio the photon momentum is effectively transferred to the ion(s) \cite{1,2}. Secondly, on the statistical level over many events, the angular distribution of the fragments is altered. This manifests itself in deviations from the predictions made within the dipole approximation.

In the present work, we focus on the often neglected first effect, namely the role of photon momentum transfer to the fragment ion(s). As introduced already in 1978 by Domcke and Cederbaum \cite{3},  the photoelectron exerts a recoil momentum onto its parent ion and, as we will demonstrate in this letter, the photon momentum adds to that photoelectron recoil, as predicted  in \cite{4}. Among the effects driven by this recoil are rotational \cite{5,6} and vibrational \cite{7,8} excitation. It manifests in high-resolution photoelectron spectra \cite{9}, Auger electron spectra \cite{6}, and can alter the electronic decay of excited states \cite{10,11}.

In our experiment, we study K-shell ionization of N$_2$ at 40~keV photon energy:
\begin{align}    
h\nu (40\text{ keV})  + \text{N}_2  \longrightarrow&  \text{N}_2^+(1s^{-1}) + e_{photo} \nonumber \\
\longrightarrow& \text{N}^+ + \text{N}^+ + e_{photo} +e_{Auger}
\end{align}
Here, the photon momentum is $k_\gamma=10.7$~a.u., the relativistic photoelectron momentum is $k_{ep}=54.9$~a.u., and the Auger electron momentum is $k_{eA}=5.0$~a.u.

The experiment has been performed at beam line ID31 of the European Synchrotron Radiation Facility (ESRF) in Grenoble, France, using a COLTRIMS (Cold Target Recoil Ion Momentum Spectroscopy) reaction microscope \cite{1,12}. A supersonic N$_2$ gas jet was crossed with the photon beam yielding a localized interaction region of $0.4\times0.1\times1.0$~mm$^3$. The photon energy was selected using a pinhole-monochromator \cite{13}, yielding a flux of $8.4\times 10^{14}$ photons/s at $\Delta E/E=1.1$\%. The sychrotron machine operated in 16-bunch mode, i.e. 5.68~MHz bunch rate, with a bunch length of 50~ps (rms). The electrons and ions where guided by a 51.7~V/cm electric field and a parallel 20.6~G magnetic field towards two position-sensitive microchannel plate detectors with delay-line anode \cite{14}. This yields $4\pi$ collection solid angle for the emitted N$^+$ ions and the Auger electron, which were measured in coincidence. From the positions-of-impact and the times-of-flight the momentum vectors of the detected particles are obtained. At 40~keV, ionization occurs via two processes, K-shell photoabsorption and Compton scattering. We distinguish these processes by inspecting the sum momentum of the ions \cite{2,15}. As  photoionization occurs, the momentum of the ions' center of mass is given by the sum of the photon momentum, the recoil of the photoelectron and of the Auger electron $\vec{k}_{cm}=\vec{k}_\gamma-\vec{k}_{ep}-\vec{k}_{eA}$, which gives a minimum value of 39.4~a.u. in the present experiment. For Compton scattering in contrast, the reaction is a binary encounter of the photon and the electron. The ionic core only has to balance the initial momentum of the bound electron \cite{16}, since the momentum of the ejected electron is balanced by the momentum change between incoming and scattered photon. For those events identified as photoionization, we obtain the momentum vector of the fast photoelectron utilizing momentum conservation.

\begin{figure}[t]
\centering
  \includegraphics{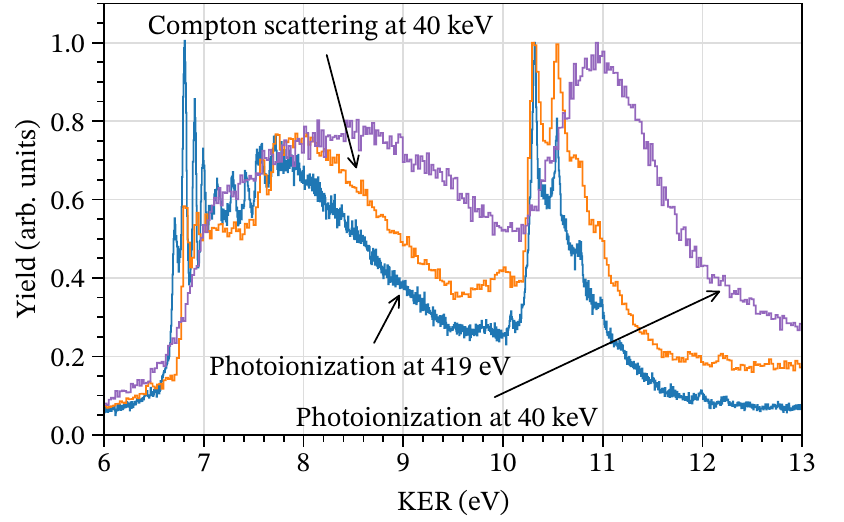}
\caption{Kinetic energy release of fragmentation of N$_2^{2+}$ into N$^+$+ N$^+$ after photoabsorption. The graph for K-shell photoionization at 419~eV photon energy (blue line) is taken from \cite{17}. Orange line: This work, N$_2^{2+}$ produced by Compton scattering at $h\nu=40$~keV. Purple line: This work, N$_2^{2+}$ produced by K-shell photoionization with photons of $h\nu=40$~keV. The data are all normalized to the highest peak, respectively.}
 \label{fig1}
\end{figure}

\mbox{Figure~1} shows the measured kinetic energy release (KER) of the $\text{N}^+/\text{N}^+$ fragments for Compton scattering and photoabsorption separately. We emphasize that both distributions are measured simultaneously. For comparison, we also show the KER for K-shell ionization at $h\nu=419$~eV taken from the literature \cite{17}. The different processes populating the dicationic states cause the difference between the distributions shown in blue and purple. For innershell photoionization the doubly charged ion is created by Auger decay while Compton scattering has an additional contribution from direct valence shell double ionization, which is expected to yield a similar KER spectrum as electron impact double ionization \cite{18}. In all cases the KER is obtained in the center of mass system of the two N$^{+}$ ions:
\begin{align} 
\text{KER}=\frac{1}{2\mu}\frac{\left|\vec{k}_{\text{N}^+_a}-\vec{k}_{\text{N}^+_b}\right|^2}{4} \, ,
\end{align}
where $\vec{k}_{\text{N}^+_{a,b}}$ are the momenta of the ions $a$ and $b$ in the laboratory system and $\mu$ is the reduced mass. We observe a significant shift of the peak centered around 10.6~eV in the KER spectrum by about 0.5~eV. This peak results from Auger decay onto the $(2\sigma_u)^{-1} (1\pi_u)^{-1}$ $^1\Pi_g$, $(1\pi_u)^{-2}$ $^1\Sigma_g^+$ and $(3\sigma_g)^{-1} (2\sigma_u)^{-1}$ $^1\Sigma_u$ states of N$_2^{2+}$ \cite{17}. The last state is responsible for the narrow peaks visible in the Compton-scattering-induced KER distribution. Ionization by Compton scattering, or K-shell ionization close to the threshold, populate mainly the ground and lowest vibrational states of the N$_2^+(1s^{-1})$ molecular ion \cite{19}. The nuclei have little kinetic energy and after the Auger decay, the final kinetic energy is mainly determined by the potential energy of the respective state within the Franck-Condon region. 

For the case of K-shell photoionization at 40~keV photon energy on the contrary, the photoelectron momentum is transferred locally to one of the atoms. For a homonuclear diatomic molecule, thus half of the momentum is imparted to the relative motion between the nuclei \cite{3,4}, where it leads to rotational and vibrational excitation. In the present case, $k_{ep}=54.9$~a.u. corresponds to a vibrational and rotational kinetic energy of 0.78~eV deposited to the internal motion of the N$_2^+(1s^{-1})$ molecular ion. This explains the significant increase of the KER visible in \mbox{Fig.~1}. The observed increase of the KER is therefore the counterpart to the decrease of the photoelectron energy due to vibrational and rotational excitation by the recoil effect \cite{3,5,6,7,8,9}.

\begin{figure}[t]
\centering
\includegraphics{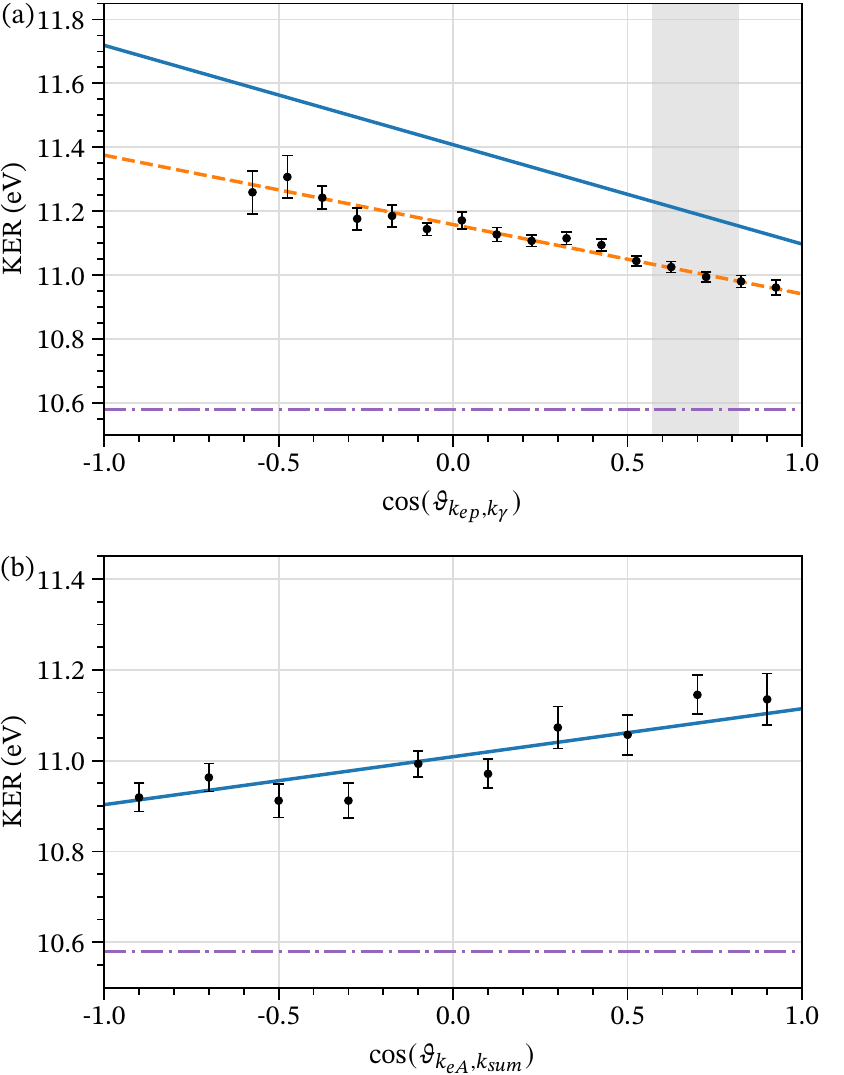}
\caption{Position of the maximum in the KER distribution between 10 and 12~eV for Compton scattering and K-shell photoionization of N$_2$ at $h\nu=40$~keV as function of $\cos{\vartheta}$, where $\vartheta$ is the angle $\vartheta_{k_{ep},k_{\gamma}}$ (panel a) or $\vartheta_{k_{eA},k_{sum}}$ (panel b), respectively. See text for explanation (including the shaded area). Dash-dotted purple lines: The KER peak position for Compton scattering (KER$_0$). Black points: Measured data for KER peak position for K-shell photoionization. Blue lines: Expectation for $\Delta\text{KER}+\text{KER}_0$, where $\Delta$KER is taken from Eq.~(\ref{eqKER}) (panel a) or Eq.~(\ref{eqKER2}) (panel b). Dashed orange line: $0.7\times\Delta\text{KER}+\text{KER}_0$ (Expectation from Eq.~(\ref{eqKER}), assuming only 70\% of the maximum recoil energy is imparted on the ionic fragments.)}
\label{fig2}
\end{figure}

In a next step we demonstrate qualitatively that the observed increase of the KER is not caused by the photoelectron recoil alone but is in part due to the photon momentum. Additionally, we show how the Auger electron momentum contributes. To experimentally prove the role of the photon momentum we inspect the KER as a function of the angle $\vartheta_{\gamma,ep}$ between the photon direction $\vec{k}_\gamma$ and the  momentum vector of the photoelectron $\vec{k}_{ep}$ (\mbox{Fig.~2(a)}). The data show a linear dependence on $\cos(\vartheta_{\gamma,ep})$. \mbox{Figure~2(a)} shows that the photon momentum changes the KER by about 0.4~eV depending on whether the photon momentum points parallel or antiparallel to the photoelectron.

To model this situation, we follow \cite{4}, assuming that  $(\vec{k}_{\gamma}-\vec{k}_{ep})/2$ is deposited into vibrational and rotational degrees of freedom N$_2^+(1s^{-1})$ molecular ion. By energy conservation the corresponding energy is distributed after the Auger decay among the ions (as additional KER) and the Auger electron. The maximum increase $\Delta$KER$_{max}$, occurring if the Auger electron energy was not influenced by the photon and photoelectron recoil momentum, is then given by
\begin{align}    
\Delta\text{KER}_{max} &= \frac{\left|\vec{k}_{\gamma}-\vec{k}_{ep} \right|^2}{4 m_{\text{N}}} \nonumber\\
&= \frac{k^2_{\gamma} + k^2_{ep}}{4m_{\text{N}}}
-\frac{k_{\gamma}k_{ep}}{2m_{\text{N}}} \cdot \cos(\vartheta_{\gamma,ep}) \ ,\label{eqKER}
\end{align} 
where $m_{\text{N}}$ is the mass of a nitrogen atom. The full blue line in \mbox{Fig.~2(a)} shows this upper bound for a shift of the KER. The dashed orange line shows that about 70\% of the maximum value is transferred to the kinetic energy release, implying that the remaining 30\% of the recoil-induced energy transfer is taken up by the Auger electron. This is in the range of what one can expect from the following simplified consideration: For diatomics, 50\% of the internal energy leads to rotational excitation. This does not couple to the Auger electron and hence is fully observed in the KER. The remaining 50\% leads to vibrational excitation \cite{9}. The distribution of this energy between the Auger electron and the KER depends on the shape of the respective potential energy surfaces. Neglecting the photon momentum would lead to a horizontal line in the figure, thus the nicely reproduced slope of angular dependence shows that the photon momentum adds to the momentum-induced energy transfer as predicted \cite{4}.

Up to now we have investigated the KER integrated over all emission directions of the Auger electron. \mbox{Fig.~2(b)} shows that the recoil momentum of the Auger electron $-\vec{k}_{eA}$ also influences the KER. Again, a linear dependence of the KER on $\cos{(\vartheta_{k_{eA},k_{sum}})}$ is found, where $\vartheta_{k_{eA},k_{sum}}$ is the angle between the Auger electron and momentum sum of photoelectron recoil and photon momentum. We have selected events where the photoelectron was emitted to an angle of 45$\pm10$~deg with respect to the photon direction, indicated by the shaded region in \mbox{Fig.~2(a)}. The corresponding recoil-induced kinetic energy change from the photoabsorption step $c\times \Delta \text{KER}_{max}$, with the above empirical value of $c=0.7$, corresponds to an effective momentum of $k_{eff}$ of 33.5~a.u. The recoil momentum imparted by the Auger electron adds to this momentum in full, leading to a total increase of the KER by
\begin{align}    
\Delta\text{KER} = \frac{k^2_{eff} + k^2_{eA}}{4m_{\text{N}}}
-\frac{k_{eff}k_{eA}}{2m_{\text{N}}} \cdot \cos(\vartheta_{k_{eA},k_{sum}}) \ , \label{eqKER2}
\end{align} 
which is shown by the full blue line in \mbox{Fig.~2(b)}. From \mbox{Eq.~(\ref{eqKER2})} and \mbox{Fig.~2(b)} it follows that in the present case, the recoil of the Auger electron modifies the KER by about 0.2~eV depending on whether the Auger electron is emitted parallel or antiparallel to the sum momentum of the photoelectron and photon. This functional dependence indicates that for N$_2$, the momentum transfer of photon and photoelectron are only partially contributing to the KER, while the recoil of the Auger electron is fully contributing. While it seems plausible that the photon momentum and the photoelectron recoil act on the same atom and thus these momenta add up, the situation may be different with respect to the Auger electron momentum. The Auger decay is not always linked to the same center as the photoemission because the vacancy can migrate across the molecule \cite{20}. For a homonuclear diatomic molecule, the hopping time is given by the energy splitting of the $g/u$ hole states \cite{19,21,22}. In N$_2$, the splitting is approximately 100~meV which corresponds to a time for the hopping from one to the other atom of $\sim$20~fs. This time is long compared to the $\sim$7~fs lifetime of the hole states, so in most cases the Auger decay takes place before the hole changes sites. A very good agreement of the present data in \mbox{Fig.~2(b)} with the classical estimation via \mbox{Eq.~(\ref{eqKER2})} suggests that the Auger electron recoil acts on the same site as the photoelectron recoil.
\begin{figure}[b]
\centering
  \includegraphics[width=1.\columnwidth]{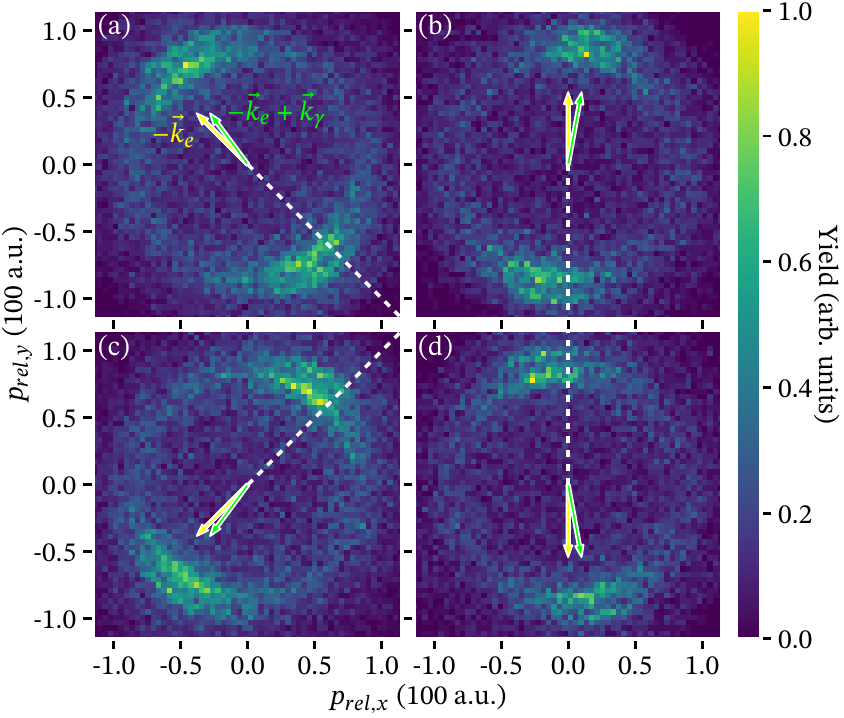}
\caption{Fragmentation of N$_{2}$ by K-shell photoionization. Displayed are the components $x$ and $y$ of the relative momentum $\vec{p}_{rel}$ (see text). The panels (a)-(d) correspond to different emission directions of the photoelectron (dashed white lines). The yellow arrows indicate the photoelectron recoil, the green arrows the sum of photoelectron recoil and the photon momentum.}
       \label{fig3}
\end{figure}

In addition to modifying the KER, the photon momentum also induces a rotational motion of the N$_2^+(1s^{-1})$ molecular ion. This is evidenced in \mbox{Fig.~3}, which shows the distribution of the relative momentum $\vec{p}_{rel}=(\vec{k}_{\text{N}^+_a}-\vec{k}_{\text{N}^+_b})/2$ of the two fragment ions. The two rings in the figure correspond to the two peaks in the KER spectrum (see \mbox{Fig.~1}). In the individual panels of \mbox{Fig.~3}, we have selected subsets of events, where the photoelectron is emitted to different directions, as indicated by the dashed lines. We find distinct maxima in the angular distributions which are {\em not} located at the angle of the photoelectron recoil, but at the angle of the sum momentum of photon momentum and photoelectron recoil. The molecules are initially randomly oriented in the gas jet and the photoionization probability at these high photon energies does not depend on the orientation of the molecule at the instant of absorption. Naively, one would therefore expect an isotropic angular distribution of the fragmentation, which is strikingly not the case in \mbox{Fig.~3}. This observed alignment of the fragmentation is produced by rotation of the N$_2^+(1s^{-1})$ molecular ion between ionization and Auger decay. The momentum transfer initializes a rotational wavepacket at the time of photoemission. This wavepacket of the originally isotropic sample evolves in time and is quenched by the Auger decay. Without this quenching, it would show revivals as they are known from non-adiabatic alignment of molecules in a strong non-resonant laser pulse \cite{23}. From these laser-based experiments, the revival time for N$_2$ is known to be 58.462~ps \cite{24}, i.e., much longer than the Auger lifetime. The laser-based experiment show abrupt alignment briefly after the laser pulse. It is this initial alignment which is seen in \mbox{Fig.~3}. A more classical perspective on the rotation already accounts for such a transient alignment along the direction of momentum transfer shortly after the kick. Qualitatively, those molecules which are oriented along the kick experience a compression or stretch, but no rotation is induced. Those molecules, however, which are aligned perpendicular to the kick receive the maximum angular momentum and hence rotate fastest. This depletes the break-up directions perpendicular to the momentum transfer. 

In the current context, the key message of \mbox{Fig.~3} is that the fragmentation maximizes along the direction of the sum momentum vectors of the photoelectron recoil and the photon momentum. The influence of the photon momentum is most strikingly seen by comparing \mbox{Fig.~3} (b) and (d). In both panels the photoelectron recoil is vertical, but points up on panel (b) and down in panel (d). The preferred breakup is, however, tilted by the photon recoil clockwise or counterclockwise; a tilt purely induced by the photon momentum.

In conclusion we have shown that in molecular photoionization the photon momentum induces rotational and vibrational motion of the molecular ion. For a homonuclear diatomic molecular, this connects to the question of core hole localization as it shows that the photon transfers its momentum locally to one or the other of two equivalent atoms. The observation clearly supports the prediction from \cite{4} and rules out a scenario where the photon momentum is split between the centers or transferred only to the center of mass of the molecule. The situation is analogous to momentum transfer to a molecule by Rutherford scattering \cite{25}, where also the momentum transfer yields a coherent superposition of vibrational states of a molecule which are created by the full local momentum being transferred coherently to one {\em and} the other center. The photon momentum plays a fully equivalent role to the recoil momentum of the photoelectron. The decisive quantity for the molecule is the sum of all momentum transfers. Following \cite{26}, one might envision that this momentum transfer could be used to not only modify the KER as in the diatomic case but for larger molecules to steer molecular dissociation pathways e.g. through conical intersections.

\begin{acknowledgments}
We thank Helena Isern and Florian Russello from beam line ID31 at ESRF for excellent support during the beam time. We acknowledge helpful discussion with Ralph P\"uttner, Kiyoshi Ueda, and Faris \mbox{Gel'mukhanov}.

This work was supported by DFG and BMBF. We acknowledge support of the theory-experiment collaboration from Deutsche Forschungsgemeinschaft via Sonderforschungsbereich 1319 (ELCH). 
\end{acknowledgments}

\end{document}